# Diversity of Intrinsic Frequency Encoding Patterns in Rat Cortical Neurons—Mechanisms and Possible Functions

Jing Kang[2], Hugh P. C. Robinson[3]*, Jianfeng Feng[1,2]*

1 Center for Computational Systems Biology, Fudan University, Shanghai, People's Republic of China, 2 Department of Computer Science, University of Warwick, Coventry, United Kingdom, 3 Department of Physiology, Development and Neuroscience, University of Cambridge, Cambridge, United Kingdom

## Abstract

Extracellular recordings of single neurons in primary and secondary somatosensory cortices of monkeys *in vivo* have shown that their firing rate can increase, decrease, or remain constant in different cells, as the external stimulus frequency increases. We observed similar intrinsic firing patterns (increasing, decreasing or constant) in rat somatosensory cortex *in vitro*, when stimulated with oscillatory input using conductance injection (dynamic clamp). The underlying mechanism of this observation is not obvious, and presents a challenge for mathematical modelling. We propose a simple principle for describing this phenomenon using a leaky integrate-and-fire model with sinusoidal input, an intrinsic oscillation and Poisson noise. Additional enhancement of the gain of encoding could be achieved by local network connections amongst diverse intrinsic response patterns. Our work sheds light on the possible cellular and network mechanisms underlying these opposing neuronal responses, which serve to enhance signal detection.

**Funding:** The work was funded by the Warwick Postgraduate Research Fellowship, Institute of Advance Study Early Career Fellowship, Code, Analysis, Repository and Modelling for e-Neuroscience (Engineering and Physical Sciences Research Council, United Kingdom), and European Commission Framework Program 6 Grant (Contract No. 012788). The funders had no role in study design, data collection and analysis, decision to publish, or preparation of the manuscript.

**Competing Interests:** The authors have declared that no competing interests exist.

* E-mail: hpcr@cam.ac.uk (HPCR); Jianfeng.Feng@warwick.ac.uk (JF)

## Introduction

In a series of experiments on somatosensory frequency discrimination in monkeys, responses of single neurons in somatosensory cortex to mechanical vibrations on the finger tips or direct oscillatory electric current stimulation were recorded [1,2,3,4]. A subset of neurons in primary (S1) and secondary (S2) somatosensory cortices showed modulations of their firing rates with the temporal input frequency ($F$). Most neurons in S1 tune with a positive slope to the input frequency, but some neurons in S2 behave in an opposite way, with a high firing rate at low stimulus frequency which is reduced at high frequency. It is unclear if these heterogeneous frequency response functions of neurons in different areas of somatosensory cortex are due to local neural network properties, receptor properties or input connectivity, or to the intrinsic integrative characteristics of single neurons.

To investigate the characteristics of single neurons, we performed whole-cell patch clamp recordings from the somas of layer 2/3 pyramidal neurons in rat somatosensory cortex *in vitro*, [5], and stimulated firing by directly injecting oscillatory artificial synaptic conductance and current into neurons through the patch-clamp pipette [6]. We found that some neurons generated a higher firing rate as stimulus frequency increased, while others showed a reduced firing rate at high frequency. We also observed a lot of frequency-insensitive neurons, which fired at a constant rate as stimulus frequencies vary. In addition, the types of neuronal responses (increasing, decreasing or constant) were affected in some cases by the mean, or offset, of stimulus intensity (see Fig. 1C, stimulus illustration). With the diversity of firing patterns observed in individual neurons in our experiments, it appears possible that the intrinsic properties of neurons can explain much of the diversity of response patterns observed *in vivo*. A reasonable goal in modeling these responses would be a simple model which could generate these different patterns as its parameters are varied.

The leaky integrate-and-fire (LIF) model is simple, analytically tractable and computationally efficient, compared with other complex biophysical models (e.g. Hodgkin-Huxley models). A number of studies have concluded that LIF neurons can not be used for simulating temporal frequency coding mechanisms at the single neuron level [7,8,9,10,11], and that the LIF model is blind in the temporal domain owing to the fact that its efferent firing rate is independent of the input temporal frequency [9]. This is true under certain circumstances, but not all. Here, we have managed to generate output firing rates in LIF models with three different patterns (increasing, decreasing or flat) as a monotonic function of the input frequency $F$, under a wider, but still biologically feasible, parameter region than considered previously. We were able to provide a simple mathematical explanation for the underlying mechanism of these three different firing patterns in the LIF model. We suggest that simple networks of these neurons could enhance the gain of frequency encoding.

## Materials and Methods

### Ethics Statement

All animals were handled in strict accordance with good animal practice as defined by the UK Home Office regulations, sacrificed according to UK Home Office approved Schedule 1 procedures,

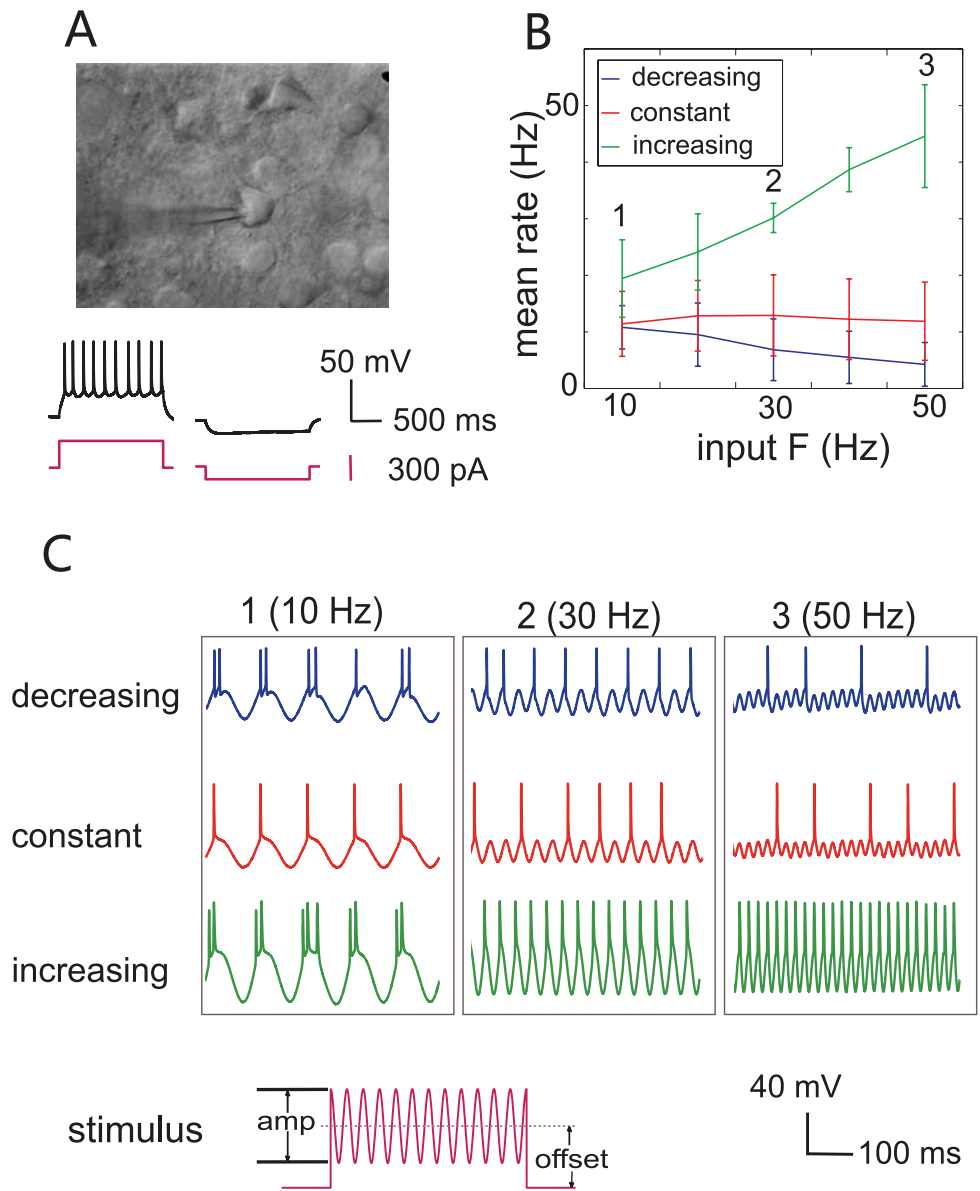

Figure 1. Experimental results. (**A**) Infrared differential interference contrast photograph of a whole-cell patch-clamp recording from a regular-spiking pyramidal neuron: stimulation and recording are carried out through the pipette on the soma. Below: recorded membrane potential (black) filtered with a Gaussian digital filter when injected constant current (pink) is 300 pA (left) and −100pA (right). (**B**) Average tuning curves of neurons when the offset values of the injected stimuli varies. The output spiking rate is a decreasing function of the input frequency (blue) when stimuli were of relatively small offset magnitude, and the neuron's firing rate was steady (red) or even increasing (green) for stimuli with larger offset. (**C**) Membrane potential with sinusoidal current injection (pink) of different frequencies of 10, 30 and 50 Hz, respectively (blue: decreasing, red: flat, and green: increasing).
doi:10.1371/journal.pone.0009608.g001

and all animal work was approved by the University of Cambridge.

### Biophysical Experiments

**Electrophysiology.** 300 μm sagittal slices of somatosensory cortex were prepared from postnatal days 7–21 Wistar rats (handled according to United Kingdom Home Office guidelines), in chilled solution composed of the following (in mM): 125 NaCl, 25 NaHCO$_3$, 2.5 KCl, 1.25 NaH$_2$PO$_4$, 2 CaCl$_2$, and 25 glucose (oxygenated with 95% O$_2$, 5% CO$_2$). Slices were held at room temperature for at least 30 min before recording and then perfused with the same solution at 32–34°C during recording. Whole-cell recordings were made from the soma of pyramidal neurons in cortical layers 2/3. Patch pipettes of 5–10 MΩ resistance were filled with a solution containing of the following (in mM): 105 K-gluconate, 30 KCl, 10 HEPES, 10 phosphocreatine, 4 ATP, and 0.3 GTP, adjusted to pH 7.35 with KOH. Current-clamp recordings were performed using a Multiclamp 700B amplifier (Molecular Devices, Union City, CA). Membrane potential, including stated reversal potential for injected conductances, was corrected afterwards for the pre-nulling of the liquid junction potential (10 mV). Series resistances were in the range of 10–20 MΩ and were measured and compensated for by the Auto Bridge Balance function of the Multiclamp 700B. Signals were filtered at 6–10 kHz (Bessel), sampled at 20 kHz with 16-bit resolution, and recorded with custom software written in C and Matlab (MathWorks, Natick, MA).

**Conductance injection.** Recorded neurons were also stimulated using conductance injection, or dynamic clamp [6,12,13]. A conductance injection amplifier (SM-1) or software running on a DSP analog board (SM-2; Cambridge Conductance, Cambridge, UK) implemented multiplication of the conductance command signal and the real-time value of the driving force, with a response time of <200ns (SM-1) or <25 μs (SM-2), to produce the current command signal. Voltage dependence of NMDA current was simulated by multiplying the command signal by an additional factor $(1+0.33[Mg^{2+}]\exp(-0.06V))^{-1}$ [14], where V is the membrane potential and $[Mg^{2+}]$ is the extracellular magnesium concentration set to 1 mM. The reversal potentials $E_{AMPA}$, $E_{NMDA}$ and $E_{GABA}$ were set to be 0, 0, and −70 mV, respectively.

**Stimulus protocol.** Randomly permuted sequences of stimuli were calculated for each combination of different values of the mean offset, amplitude and frequency of the sinusoidal input (Fig. 1C, stimulus), either as injected positive current or excitatory conductance, in order to obviate the effects of any progressive adaptation to monotonic changes of any single parameter. Individual sweeps consisted of 2 s of stimulus, with data from the initial 200 ms [15,16] discarded to eliminate transient onset responses. A 15 second interval between sweeps was allowed for recovery. A small hyperpolarizing holding current (<50 pA) was applied if necessary to ensure a fixed resting potential between sweeps, usually between −65 to −75 mV. Step current injections from −100pA gradually increasing with a step size of 100pA were applied at the beginning, in order to determine the neuron's capacity to stimulus intensity and assess the feasible range of the current and conductance injection within which neurons were able to generate action potentials.

**Data analysis.** The occurrence of spikes was defined by a positive crossing of a threshold potential, usually −40mV. Spike rate was calculated as the number of occurrence of spikes over the total time period (1.8 s). Of 23 cortical neurons recorded, 11 regular-spiking (RS) cells were selected for detailed analysis, whose average membrane time constant was 22.7±8.5 ms.

### Mathematical Modelling

**Single neuron model.** Mathematical modeling was based on a simple but analytically tractable model of a spiking neuron — the integrate-and-fire model. Action potentials are generated by a threshold process. Let $v(t)$ be the membrane potential of the neuron, $V_\theta$ the threshold, and $V_{rest}$ the resting potential. Suppose $V_\theta > V_{rest}$, and when $v(t) < V_\theta$, the leaky integrate-and-fire model has the form

$$\begin{cases} dv(t) = -\frac{v(t) - V_{rest}}{\gamma} dt + dI_{syn}(t), \\ v(0) = V_{rest} \end{cases} \quad (1)$$

where $\gamma$ is the decay time constant, $I_{syn}(t)$ is the synaptic input defined by

$$dI_{syn}(t) = \mu(t)dt + \sigma(t)dB_t, \quad \text{where } \mu(t) \geq 0, \sigma(t) \geq 0,$$

and $B_t$ is standard Brownian motion.

The synaptic current $I_{syn}$ is composed of two terms: a deterministic driving force $\gamma\mu$ that depolarizes the cell to fire, and a perturbing noise term $\gamma\sigma$. We assume that a neuron receives synaptic inputs from $N_s$ active synapses, each sending Poisson EPSPs (excitatory post-synaptic potentials) inputs to the neuron with rate

$$\lambda_E(t) = \frac{a}{2}(1 + \cos(2\pi Ft)),$$

where $a$ (magnitude), $F$ (temporal frequency) are both constant, and $t$ is the time. More specifically, $\lambda(t) = \lambda_E(t)N_s$ is the input rate, and the Poisson process inputs are defined by $\mu(t) = \lambda(t)$ and $\sigma^2(t) = \lambda(t)$ [9]. A refractory period $t_{ref}$ from 1 to 5 ms is also introduced in the model, matching the observation of membrane potentials in the experiment. The input temporal frequency $F$ is confined within the range of 1 to 50 Hz, consistent with feasible biological frequencies [1,17]. In this paper we concentrate on the mean output firing rate with respect to different input frequencies.

**Recurrent excitatory network neurons.** In a neural network of size N, we assume that neuron i is connected to neuron j by a connection weight $w_{i,j}$ (drawn randomly from a standard normal distribution), i, j = 1,..., N, and $w_{i,i} = 0$. Assume that the ith neuron generates a spike at time $t_{ip}$, $1 \leq p \leq k_i$, where $k_i$ is the number of spikes that the ith neuron generated within a certain time. The ith neuron receives the sensory synaptic current input $I_{i,syn}(t)$ and local synaptic input from the other N−1 neurons. The behavior of the membrane potential $v_i(t)$ of the ith neuron at time t is then given by

$$dv_i(t) = -\frac{v_i(t) - V_{rest}}{\gamma} dt + dI_{i,syn}(t) + dt \sum_{j=1, j\neq i}^{N} \sum_{t_{ip} < t_{jq} < t} w_{j,i} \delta(t - t_{jq}).$$

When neuron $i$ fires, it induces synaptic current in its connected neurons in the network, and their membrane potential will either increase or decrease in proportion to the synaptic connection weight, depending on the type of the synaptic input (EPSP, IPSP).

## Results

### Experiment

We carried out experiments to record from neurons in acutely-isolated slices of somatosensory cortex of the rat. Although in these conditions, the normal peripheral afferent pathways are of course removed, the intrinsic spike-generating properties of neurons are believed to be largely intact, and can be investigated under controlled conditions. Regular-spiking neurons were selected by their pyramidal appearance and their membrane potential responses to constant step current stimuli (Fig. 1A). 113 sets of stable recordings suitable for analysis in different conditions of stimulus amplitude, offset and frequency in 11 neurons were obtained. Of these, 21 out of 113 recordings showed an increasing firing rate as the input frequency increased from 10 Hz to 50 Hz, 28 recordings showed a decreasing firing rate with respect to the stimulus frequency, and the remaining 64 recordings showed no significant changes of firing rate as input frequency was varied. The averaged response rates of each category of firing pattern as a function of input frequency are plotted in Fig. 1B (mean ± STD). We found that when the stimulus offset was relatively small in comparison to the neuronal input conductance (see Methods), some neurons were able to fire at low frequency but decreased their response rate as the stimulus frequency increased (Fig. 1B, blue line). In other recordings, neurons fired in proportion to the stimulus frequency, with a positive slope, when the stimulus offset was relatively high (Fig. 1B, green line). A pattern in which firing rate remained constant as stimulus frequency varied was commonly observed as well (Fig. 1B, red line). Fig. 1C shows examples of the recorded membrane voltage in different types of response patterns at 10 Hz, 30 Hz and 50 Hz stimulus frequencies.

In some cases, individual neurons could shift from a decreasing pattern of response (with increasing stimulus frequency) at low stimulus offset amplitude, to an increasing pattern, at higher offset amplitude. This undoubtedly reflects the relationship between the threshold, the timescale of subthreshold leaky integration, and stimulus offset amplitude, which is clearly an important feature for determining the type of response. Such a shift in response pattern may not be physiologically significant, if the sensory synaptic input is in a restricted range of amplitudes.

### Single Neuron Simulation

We used an integrate-and-fire model for the simulation, studying the neuronal responses to the deterministic and stochastic (Poisson noise) oscillatory current stimuli. Every simulation was run 1000 times for the stochastic Poisson inputs. The simulation time for each neuron was 1000 ms. The modelling parameter values are $V_\theta = 20$ mV, $V_{rest} = 0$ mV, and $\mathcal{N}_s = 100$, unless otherwise specified. We choose parameter values in agreement with our experimental data from the single cell recordings and with data from the literature [9,18].

**Constant firing rate.** The LIF model had a constant firing rate when the parameters satisfied $C\gamma > V_\theta$, where $C = a\,\mathcal{N}_s / 2$ (see Appendix S1 for details). With the parameters $\gamma = 20$ ms, $a = 20.5$ Hz, and the refractory period $t_{ref} = 5$ ms, the firing rate was essentially invariant with respect to the input frequency, no matter if noise is applied in the model (Fig. 2A, purple) or not (Fig. 2A, black), consistent with the biological data (Fig. 1B, red line). Although the tuning curve for spike rate showed a local peak at around 20 Hz (compare to fluctuations in the flat experimental response pattern, Fig. 1B), this smoothed when Poisson noise is added. Membrane potential responses are plotted in Fig. 2B for three different input frequency values $F = 10$ (top), 30 (middle), and 50 (bottom) Hz, and for both deterministic and noisy input. A constant efferent firing rate means that no information about the temporal input frequency $F$ is contained in the output firing rate. Hence, by reading the efferent firing rate alone, it is impossible to perform discrimination tasks between various input frequencies, for this kind of response pattern.

One hypothesis to explain this phenomenon is that the model averages out the information in time domain. This was proposed by Feng and Brown (2004) to explain why the integrate-and-fire model neuron is insensitive to the input temporal frequency in the. They examined low input rates varying from 1 to 10 Hz, and found that the output firing rate remained a constant. When $F$ is

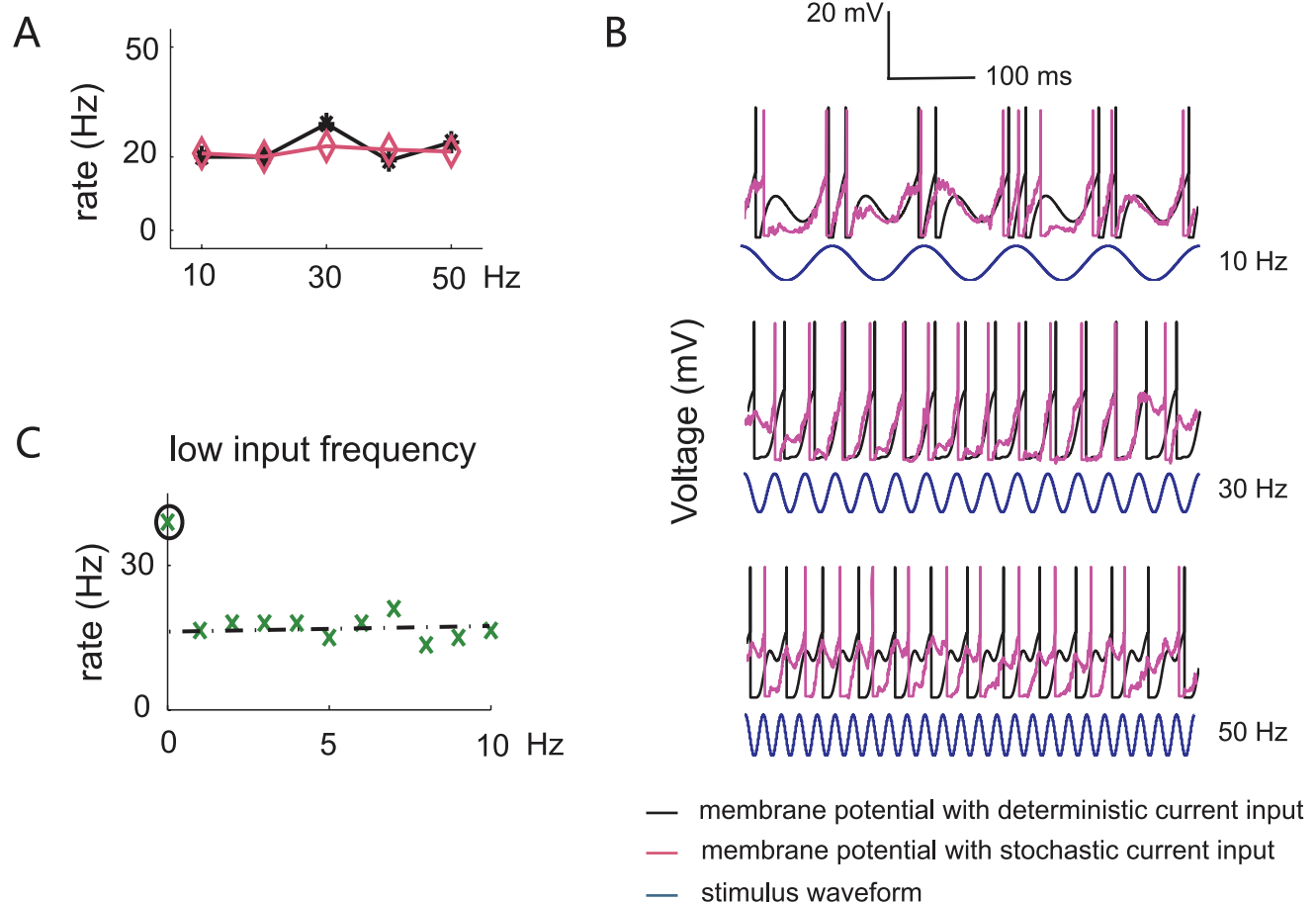

**Figure 2. Simulation results for single neurons with flat output firing rates.** (**A**) Tuning curve of a simulated neuron with parameter values: $a = 20.5$, $\gamma = 20$ ms, and $t_{ref} = 5$ ms, with (pink) or without (black) noise. (**B**) Membrane potential responses of the integrate-and-fire model to different input frequencies (top: $F = 10$ Hz; middle: $F = 30$ Hz; bottom: $F = 50$ Hz). (**C**) Except at $F = 0$ Hz, the resting output firing rate remains constant when $F$ is close to zero.
doi:10.1371/journal.pone.0009608.g002

high, the firing rate of the neuron model is given by

$$\lambda(t) = \frac{a}{2} \lim_{T \to \infty} \left[ 1 + \int_0^T \frac{\cos(2\pi Ft)dt}{T} \right] = \frac{a}{2}.$$

This finding is reproduced here in Fig. 2C. Another interesting phenomenon is that there is a sudden decrement in the value of efferent firing rates from $F=0$ to $F>0$ (Fig. 2C), which means that the integrate-and-fire model can easily detect whether or not an oscillating signal is present, but cannot tell how fast the period of the signal is.

**Decreasing efferent firing rate.** When C $\gamma<V_\theta$, the neuronal efferent firing rate is a decreasing function of the stimulus frequency (Fig. 3A). The parameter values used here are $\gamma = 20$ ms, a = 16.8 and $t_{ref} = 1$ ms. The neuron stops firing when the input frequency reaches the critical value F* = 41 Hz (see Appendix S1, Eq. 4 for detailed calculation). Membrane potential responses and input synaptic current are shown in Fig. 3B at three different frequencies (F = 10 (top), 30 (middle), and 50 (bottom) Hz), for deterministic and stochastic input. This clearly illustrates that firing rate decreases with increasing input frequency.

To further elucidate the cause of this decreasing relationship, we plotted neuronal response rate at three different stimulus amplitudes $a$ (16.8, 15 and 14) for deterministic input (Fig. 3C, top) and stochastic input (Fig. 3C, bottom). Before the neuron's firing is quenched (when $F>F^*$), even though the output firing rate is increasing over some segments of the input range (due to the phase locking under this parameter region, see Appendix S2 for a detailed explanation), its overall trend is decreasing. When Poisson noise is added, the relationship is smoothed, giving an almost monotonically decreasing trend.

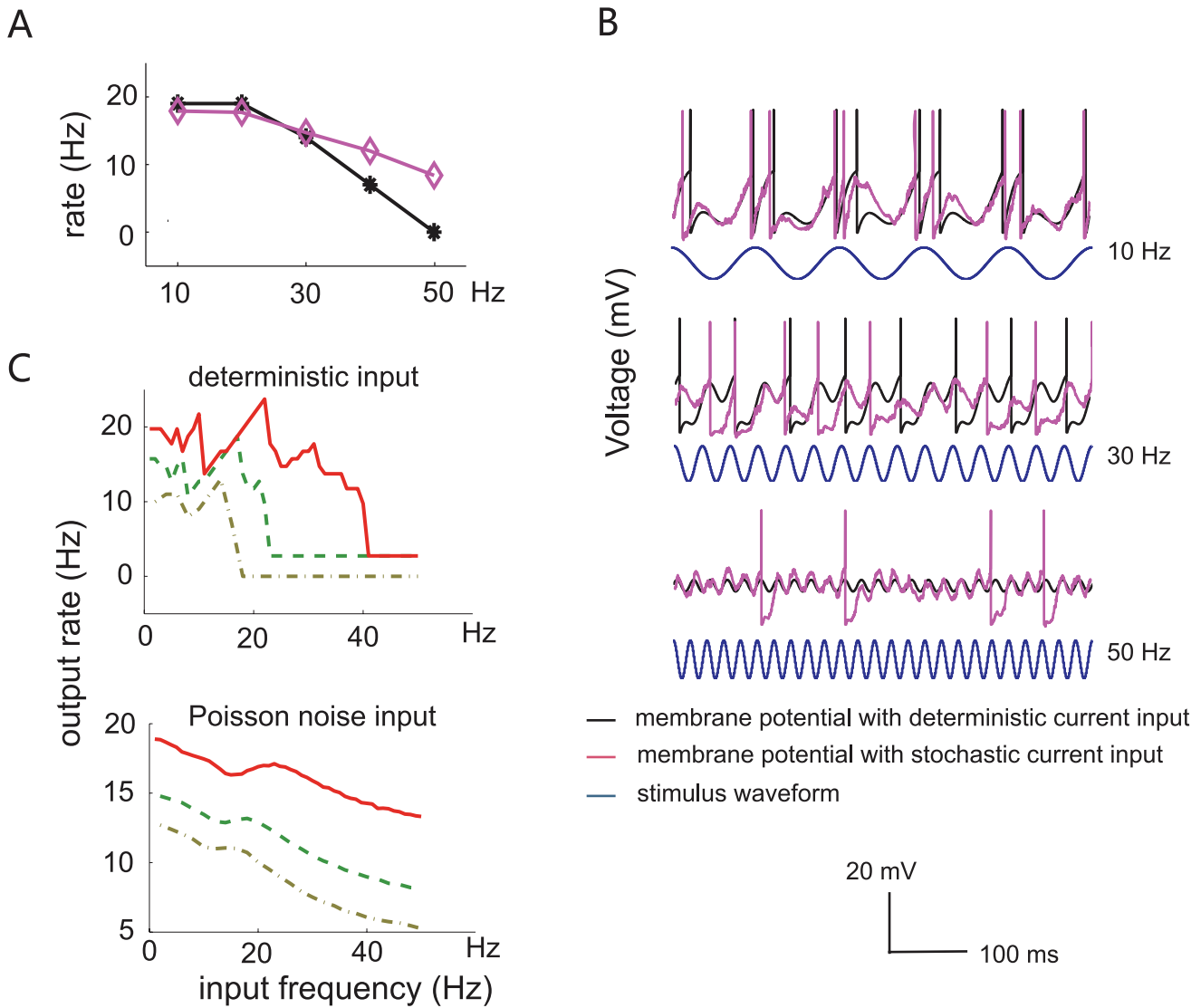

**Figure 3. Simulation results for neurons with decreasing output firing rate.** (**A**) Simulated output firing rate versus the input frequency at 10, 20, 30, 40 and 50 Hz with (black) and without (pink) noise, when parameters are: a = 16.8, $\gamma$ = 20 ms, and $t_{refr}$ = 1 ms, over the range up to 50 Hz. (**B**) Membrane potential responses of the integrate-and-fire model at different input frequencies (top: F = 10 Hz; middle: F = 30 Hz; bottom: F = 50 Hz) when noise was absent (black) or present (purple). (**C**) Input-output relation of the output firing rate versus the input frequency from 1–50 Hz continuously with deterministic input (top panel) and Poisson noise (bottom panel). The parameters are a = 16.8 (red solid line), 15 (green dash line), and 14 (brown dotted line). Here, $\gamma$ = 20 ms, and the neuronal response rates for Poisson noise were averaged over 1000 runs.
doi:10.1371/journal.pone.0009608.g003

**Increasing firing rate.** To generate an increasing spiking rate with respect to the stimulus frequency, a subthreshold intrinsic oscillation $k(\cos(2\pi\omega_0 t)+1)$ is added to the model, where k and $\omega_0$ are constant. The peak response rate is reached at the value where the input frequency F fully resonates with the intrinsic neuronal frequency $\omega_0$. Neglecting the noise term in the system, the model is fully defined by

$$\begin{cases} \dfrac{dv}{dt} = -\dfrac{v(t)-V_{rest}}{\gamma} + \dfrac{dI_{syn}(t)}{dt} + k(\cos(2\pi\omega_0 t)+1) \\ v(0) = V_{rest} \end{cases}.$$

When $\gamma = 9$ ms, $a = 10$, $t_{ref} = 5$ ms and $k = 1.5$, the efferent firing rate is an increasing function of the temporal input frequency $F$. The maximal response rate is reached at $F = \omega_0 = 50$ Hz (Fig. 4A, black). When Poisson noise is added, the tuning curve becomes smoothly monotonically increasing (Fig. 4A, purple). Fig. 4B shows example membrane potential trajectories for different input frequency values ($F = 10, 30, 50$ Hz).

The goal of our mathematical modeling is to seek a *simplest* or *minimal* mechanism to mimic the three response patterns shown by biological neurons, rather than giving a detailed biophysical model of spike generation. The simplest LIF model without any modification is capable of generating constant and decreasing firing patterns in terms of input frequency. However, in order to make the spiking rate an increasing function of input frequency, the minimal addition to the model is to include an intrinsic oscillation, where firing increases up to a peak value when the external frequency resonates with the intrinsic oscillatory frequency.

### Mechanism of Various Spiking Patterns

We next analyze the underlying mechanism of these three different response patterns. The reason for these distinct patterns can be understood in the relative location of the limit cycle of the neuronal dynamics, defined by the sinusoidal input and the "integrate" part of the integrate-and-fire model (in the absence of the spiking mechanism), and the threshold (Fig. 5 and appendix S1). A limit cycle is obtained when there is no threshold operation applied to the membrane potential, so that the three-dimensional dynamical system of the membrane potential is attracted to its stable trajectory (Appendix S1, Eq. 5).

When the limit cycle is located totally above or below the threshold, the output firing rates are all constant. In fact, when the limit cycle is below the value of the threshold, the neuron's firing rate would be zero. This is because when the membrane potential reaches the limit cycle, it will stay there forever, never crossing threshold. If the limit cycle lies above the threshold, the output firing rate is roughly constant. This is the case for flat efferent firing rate ($\gamma = 20.5$ ms). The limit circle is located above the threshold (Fig. 5 left column), and consequently, the membrane potential $v(t)$ reaches the threshold before it reaches the limit circle and is then reset to the initial value. Thus, the input frequency $F$ cannot influence the system's firing rate much. As a result, whenever the limit cycle is located completely below or above the threshold, the output firing rate is constant (zero for subthreshold case) and does not contain any information about the input frequency. An additional point is that the limit cycle is more tilted for small values of $F$ ($= 10$ Hz) than for big values (50 Hz) (see Fig. 5 left column and Appendix S1 Eq. 6 for detailed analysis).

When the limit circle intersects with the threshold (Fig. 5, middle column), the output spiking rate decreases until the input frequency $F$ increases to the critical frequency $F^*$ (see Appendix S1, Eq. 4), when the firing rate becomes zero. This pattern occurs because the limit cycle becomes flatter as $F$ goes up, causing slower spiking, but eventually comes to lie completely below the threshold, whereupon the neuron stops firing.

An alternative explanation for the constant and decreasing output firing rate versus input frequency comes from the view of phase mapping, the mapping from the phase of forcing at one spike to the next [18]. Keener et al (1981) classified the LIF neuron responses to oscillatory input into three parameter regions (see Appendix S2 for an explanation of their work and the relationship with our model). The parameter values used in our model fall into region II (piecewise phase locking) and region III (firing termination) in Keener's paper. When Poisson noise is added, the discontinuities due to the piece-wise phase locking pattern in neuronal firing rate are smoothed out, and the response curves show a consistently flat or decreasing trend versus input frequency.

Introducing an intrinsic oscillation in the neuron model is necessary to generate an increasing output spiking pattern as input frequency increases. The right column of Fig. 5 shows the limit cycle with an intrinsic oscillation term (at 50 Hz) at input frequency $F = 10$ and $F = 50$ Hz. The threshold value lies between the maximum and minimum values on the limit cycle.

### Gain Enhancement

**Network neurons.** Even though the single neuron is sophisticated enough to generate different patterns of firing rate with various input frequencies, a population of neurons connected with each other in a network can perform much better than a single neuron. We assume that neurons in the network are identical, receive the same input, and are connected with each other by excitatory synapses [19] (see Fig. 6A for an illustration of the network structure). The LIF parameters used in the network neurons are the same as for single neurons, and their connection weights are assigned randomly from a standard normal distribution. The simulation results showed that a neural network's spiking rates at different input frequencies were more distinguishable than that of a single neuron. Fig. 6 shows the decreasing and increasing firing rate patterns of the integrate-and-fire model network with random connection weights of various sizes ($N = 1, 25$, and 40 for decreasing responses; $N = 1$ and 10 for increasing responses). It can be seen that the discrimination ability of the network is better than that of a single neuron since the difference of spike rates between two frequencies in neural network is much bigger than for a single neuron, for networks of both decreasing and increasing response patterns. Neural networks with non-identical neurons whose threshold values vary ($V_\theta$ uniformly distributed within range [19.5, 20.5] mV) were also simulated, to test for the robustness of the network model, and no significant differences were found compared to identical-neuron networks (data not shown).

### Discussion

We measured experimentally the discrimination ability of single somatosensory neurons *in vitro* for temporal input frequency, in terms of their mean response rate. The LIF model was used to reproduce the results by simulation, allowing us to propose a simple underlying dynamical basis for the various patterns of neuronal responses. Our work sheds light on the possible cellular and network mechanisms of the heterogeneous frequency tuning of somatosensory cortical neurons.

### Experimental Responses

In [1,17], it was found that some neurons in the somatosensory S2 area have a lower firing rate (around 20 Hz) for high-frequency stimuli compared to the strong responses (around 40 Hz) they show to the low-frequency stimulus, but high stimulus frequencies did not completely stop the neurons from firing. However, in the present experiments, we observed a progressive reduction in firing rate with increasing input frequency, and in many instances, quenching of firing at relative high frequency. This dissimilarity might be from the differences between *in vitro* and *in vivo* conditions, affecting the intrinsic spike-generating dynamics of neurons, but could also reflect receptor and synaptic adaptation, and locally-recruited cortical inhibition.

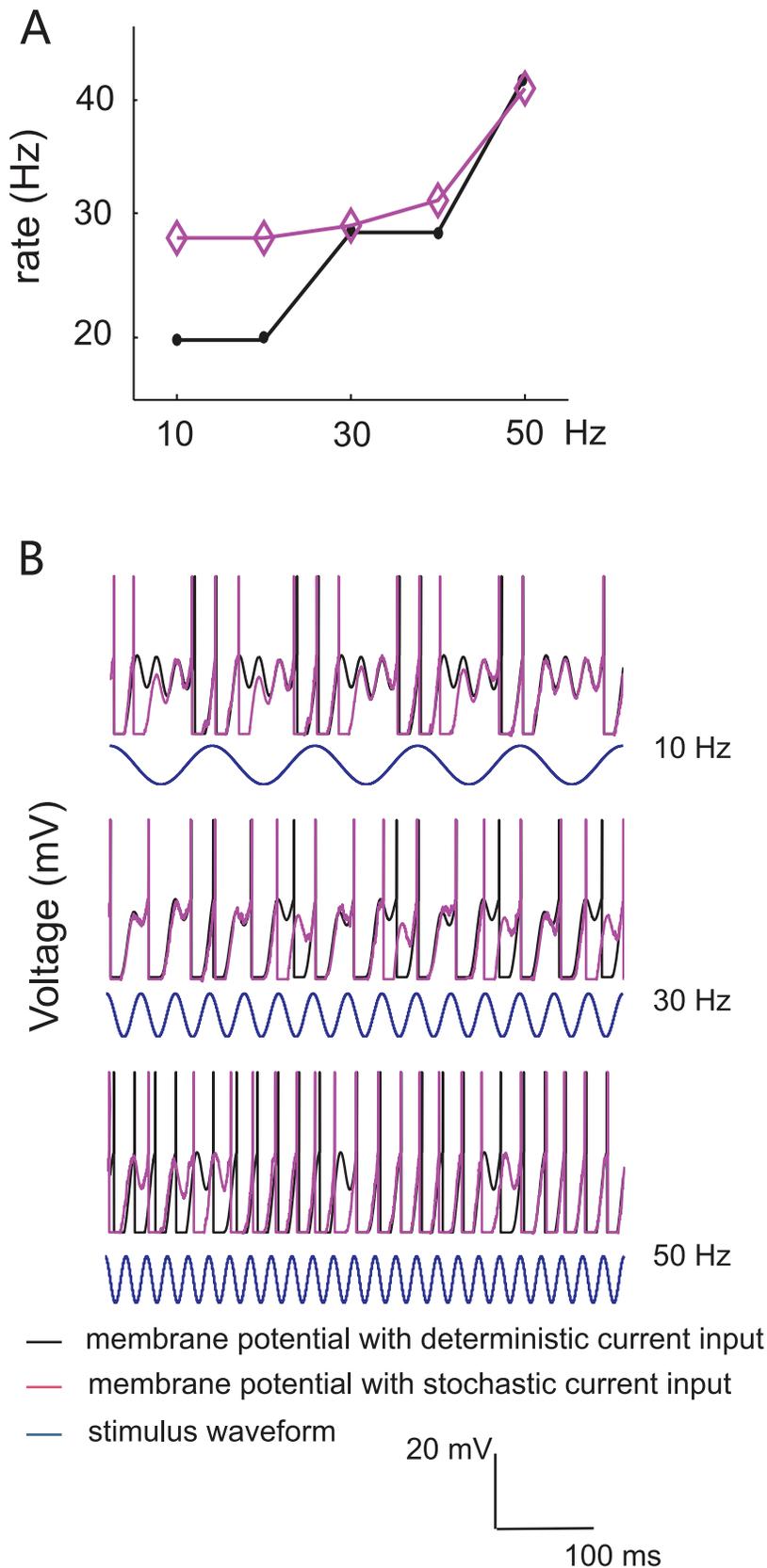

**Figure 4. Simulation of a neuron with increasing output firing rate, when an additional subthreshold intrinsic oscillation ($\omega_0 = 0.05$, $k = 1.5$) is included in the dynamical system.** Other parameter values used for modeling are $a = 10$, $\gamma = 9$ ms, and $t_{ref} = 10$ ms. (**A**) Response frequency rises as input frequency increases. (**B**) Membrane potential of the integrate-and-fire model with different values of input frequencies (top: $F = 10$ Hz; middle: $F = 30$ Hz; bottom: $F = 50$ Hz). It is seen that the neuron is more active at high frequency, and has a monotonically increasing firing pattern.
doi:10.1371/journal.pone.0009608.g004

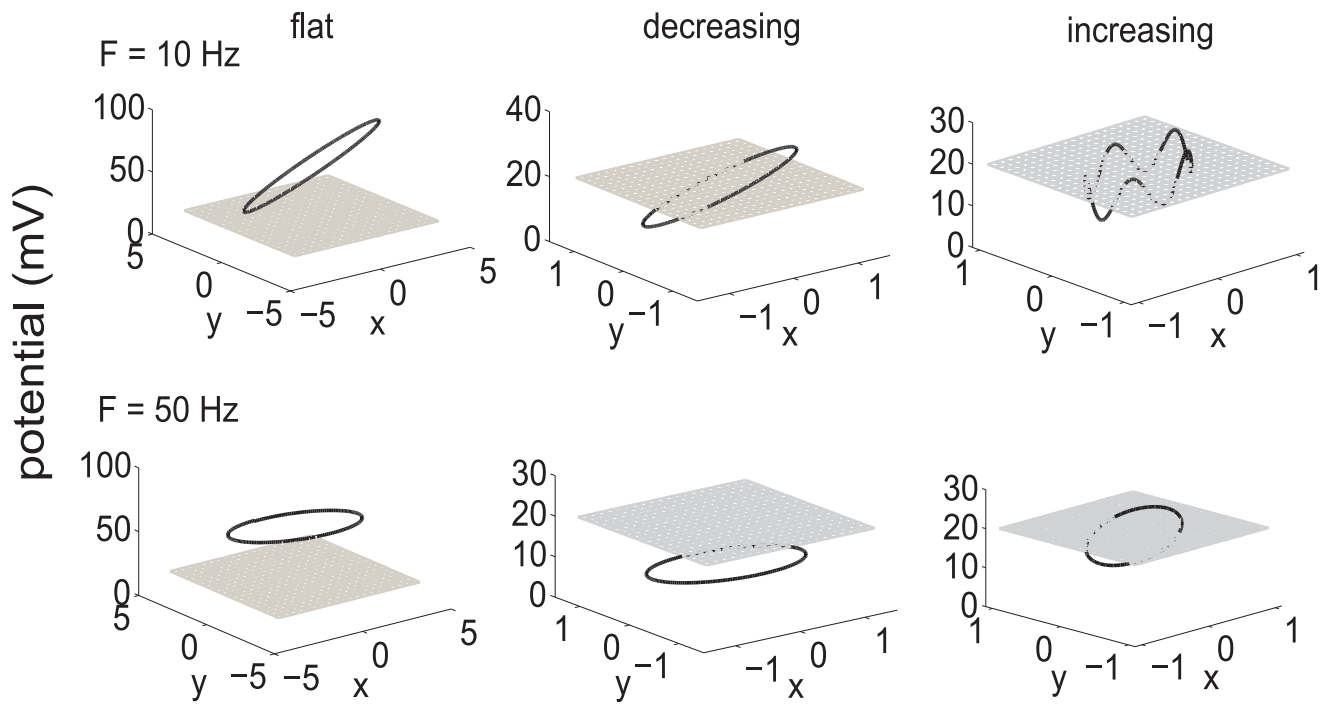

**Figure 5. Limit cycle plots for the flat, decreasing and increasing firing patterns, when no threshold is applied in the neuron model.**
A detailed explanation of this autonomous dynamical system can be found in the Appendix S1, where x axis represents $x = C\cos(2\pi Ft)$, and y axis is $y = C\sin(2\pi Ft)$. The degree of tilt when $F = 10$ Hz is much larger than when $F = 50$ Hz. Threshold value is represented by the grey grid square.
doi:10.1371/journal.pone.0009608.g005

Nevertheless, the quenching of firing observed experimentally is consistent with the behaviour of the LIF neuronal model. Experimentally, neurons decreased their firing rate versus the input frequency only when the injected current offset was close to the minimal feasible range of stimuli, for which generation of spikes was guaranteed. This minimal feasible range of stimuli of real neurons corresponds to the mathematical explanation of intersection (see Fig. 5, middle column and Appendix S1 for details) between the threshold value and the limit cycle of the dynamics. Biological neurons appeared to have a constant or increasing response versus input frequency when the oscillatory stimulus offset is in the middle range of the feasible stimuli intensity, and this is consistent with our model parameter region as well.

To compare how accurately experimental and modelled neuronal responses encode stimulus frequency, we compare them using neurometric performance curves, as shown in Fig. 7. A detailed description of the generation of neurometric curves can be found in [17]. In Fig. 7, neurometric curves were generated by plotting the percentage of each recorded data at different comparison stimulus frequencies ($F = 10, 20, 30, 40$ and $50$ Hz) for which the comparison frequency was called higher than the base frequency, which was fixed at 30 Hz (because it is the middle point of the stimulus frequency range), as a function of the comparison frequency. Points near 0% or 100%, where the base frequency and comparison frequency are very different, correspond to easy discriminations, whereas points near 50% correspond to difficult discriminations. Both for the increasing and decreasing neural responses, the neurometric functions of the modeling were considerably better than the experimental data.

### Intrinsic Oscillations in Increasing Response Patterns

An intrinsic oscillation in the frequency range of 40 to 50 Hz of pyramidal neurons, as is predicted to be required by the model to generate increasing responses, has not been clearly described in the literature. However, it should be pointed out that what is predicted is not necessarily a detectable *subthreshold* oscillation of membrane potential, but an intrinsic oscillation within the suprathreshold spiking dynamics which interacts with and resonates with an "integrate-and-fire" like component of the dynamics. A strong candidate for this would be recruitment of the local fast-spiking inhibitory interneuron network, and its feedback on the recorded pyramidal neuron [20,21,22]. Thus, it would be of interest in further studies to characterize input frequency responses in the presence of synaptic blockers of glutamate and GABA receptors to disconnect this component of the network.

### Biological Function

In the nervous system, encoding and decoding is accomplished at a system level rather than at a single neuron level. Network neurons gain an advantage in generating more distinguishable efferent spike rates at different input frequency levels, by the connectivity of the neurons in the network: one neuron's action potential will contribute to other neurons' membrane potential in proportion to the connection weight. As a result, the output firing rate of the whole neural network is boosted by positive feedback over the output rate of an individual neuron.

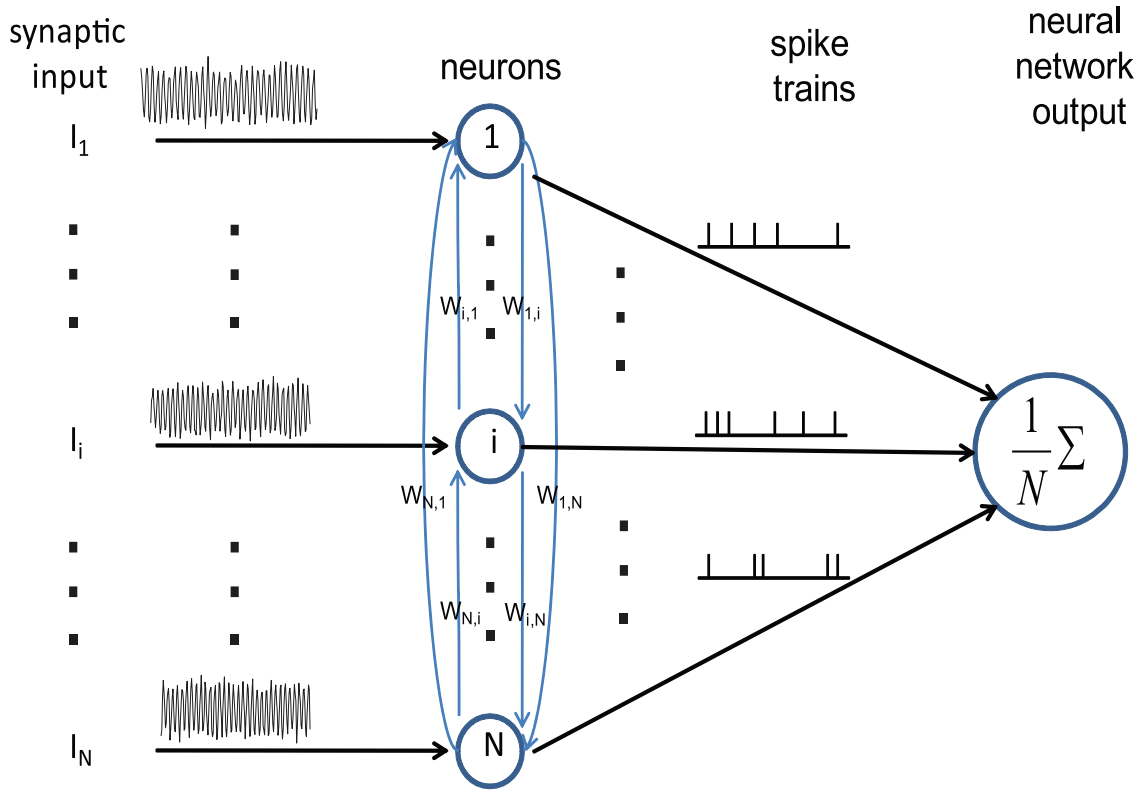

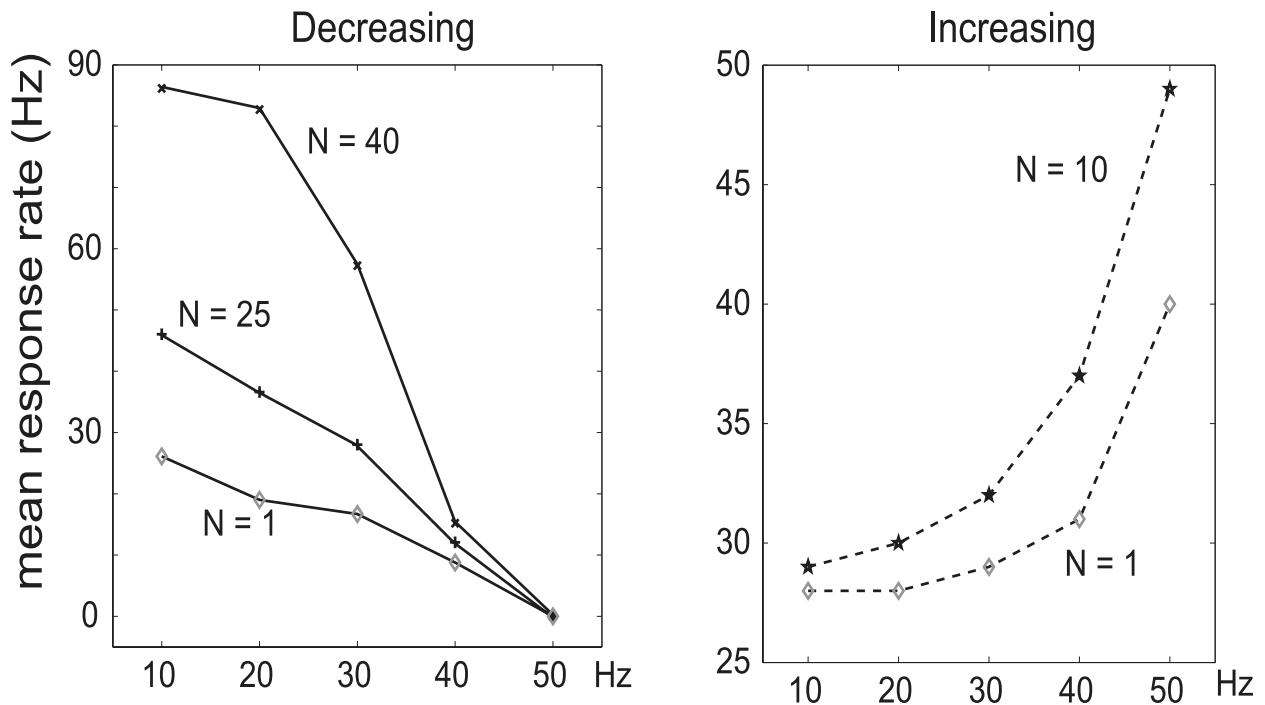

**Figure 6. Network enhancement of response gain.** (**A**) Illustration of the structure of the neural network. (**B**) The average firing rate of network neurons, for both the increasing (right) and decreasing (left) patterns. The network neurons revealed a bigger difference between the minimum and maximum firing rate than that of single neurons, both for increasing and decreasing patterns. The larger the neural network size, the more significant was the difference among neural response rates at various input frequencies. The connection weights among neurons in the network are randomly generated following a normal distribution.
doi:10.1371/journal.pone.0009608.g006

What is the biological function of these different, opposed neural tunings, especially the opposite tuning in the cortex? In experiments on electric fish [23,24,25], opposite types (increasing and decreasing) of frequency responses of electroreceptor cells in the lateral line organs have also been observed, and it was shown that electric fish recognize objects by centrally comparing the responses from these two different types of receptor cells. In [17], it is shown that cortical networks can enhance the neural representation of features from complementary populations of cells with positive and negative response slopes. Thus, gain could in principle be further increased by neurons which integrate the outputs of excitatory and inhibitory subnetworks.

### Other Possible Neural Models

The leaky integrate-and-fire model is not the only model that is able to decode the input frequency from its efferent firing rate, although using LIF alone we can account for many biological phenomena, see for example [26]. One of the other possible forms is the quadratic integrate-and-fire model [27] that we have found can make the output firing rate a decreasing function of the input frequency (data not shown). The principle is similar to what we analyzed in the leaky integrate-and-fire neuron. A more biophysically-realistic neuron model is the Hodgkin-Huxley (HH) model [28]. According to Feng and Brown (2001), the tuning curve has two maximum points and one minimum point, but it is not possible to uniquely read out the input temporal frequency [9]. The reason why the Hodgkin-Huxley model is able to generate an increasing pattern at low input frequencies is believed to be that the HH model itself contains an intrinsic subthreshold oscillation with a defined frequency, which makes it possible to generate two peaks at 60 Hz and 120 Hz, respectively, for the standard Hodgkin-Huxley model (refer to the Appendix of [9] for detailed equations and parameters).

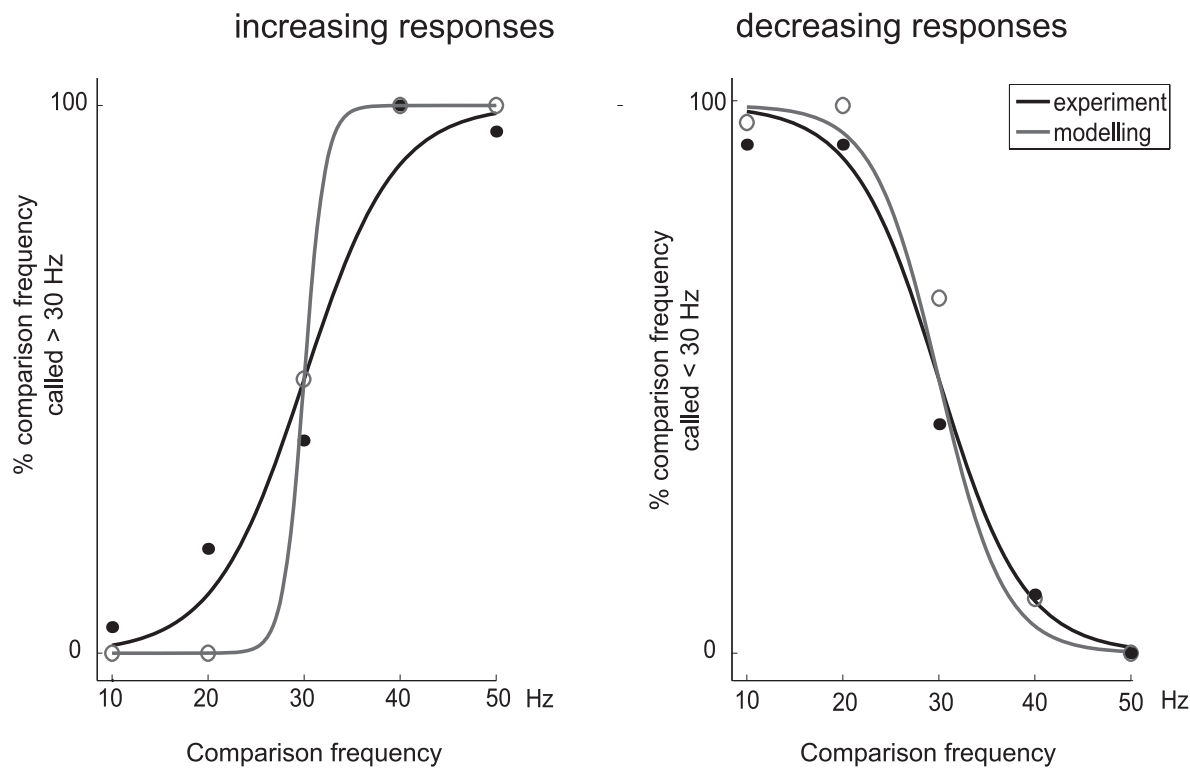

**Figure 7. Neurometric functions for the increasing and decreasing responses of the experimental recordings and the mathematical models. Left:** For neuronal response with a positive slope. Continuous curves are sigmoidal fits ($\chi^2$, $p<0.001$) to the data points for the five comparison stimulus frequencies (10, 20, 30, 40 and 50 Hz) paired with a reference stimulus frequency fixed at 30 Hz. y axis is equivalent to the probability that the comparison frequency is judged higher than the reference frequency (30 Hz). Gray line is neurometric function of experimental data; black line is of modeling data. **Right:** Same format as panel on the left, but for neuronal responses with a negative slope.
doi:10.1371/journal.pone.0009608.g007

## Supporting Information

**Appendix S1** Supplementary material for the main article.
Found at: doi:10.1371/journal.pone.0009608.s001 (0.79 MB RTF)

**Appendix S2** Comparison with a circle mapping model.
Found at: doi:10.1371/journal.pone.0009608.s002 (0.11 MB RTF)

## Acknowledgments

We thank Nam-Kyung Kim and Ben Fisher for their help in experiments, Yulia Timofeeva and Xuejuan Zhang for their valuable comments on the mathematical modeling, Jianhua Wu for his criticism on the manuscript, and Martin Pumphrey for editorial assistance.

## Author Contributions

Conceived and designed the experiments: JK HPCR JF. Performed the experiments: JK HPCR. Analyzed the data: JK. Contributed reagents/materials/analysis tools: JK HPCR JF. Wrote the paper: JK HPCR JF.